\documentclass[sigconf]{acmart}
\usepackage{graphicx}
\usepackage{caption}
\usepackage{subcaption}
\usepackage{color,soul}
\usepackage{multirow}
\usepackage{flushend}
\usepackage{array}
\usepackage{enumitem}
\usepackage{adjustbox}

\newcommand{\mysystem}{{CoQEx}}
\newcommand{\systemurl}{https://github.com/ghoshs/CoQEx}
\renewcommand{\paragraph}[1]{\smallskip\noindent\textbf{#1.\mbox{\ \ }}}

\newcommand{\squishlist}{
 \begin{list}{$\bullet$}
  { \setlength{\itemsep}{0pt}
     \setlength{\parsep}{1pt}
     \setlength{\topsep}{1pt}
     \setlength{\partopsep}{0pt}
     \setlength{\leftmargin}{1.5em}
     \setlength{\labelwidth}{1em}
     \setlength{\labelsep}{0.5em} } }
\newcommand{\squishend}{\end{list}}

\copyrightyear{2022} 
\acmYear{2022} 
\setcopyright{rightsretained} 

\acmConference[SIGIR '22]{Proceedings of the 45th International ACM SIGIR Conference on Research and Development in Information Retrieval}{July 11--15, 2022}{Madrid, Spain.}
\acmBooktitle{Proceedings of the 45th Int'l ACM SIGIR Conference on Research and Development in Information Retrieval (SIGIR '22), July 11--15, 2022, Madrid, Spain}
\acmDOI{10.1145/3477495.3531870}
\acmISBN{978-1-4503-8732-3/22/07}

\usepackage{etoolbox}
\makeatletter
\patchcmd{\maketitle}{\@copyrightpermission}{
  \begin{minipage}{0.3\columnwidth}
     \href{http://creativecommons.org/licenses/by/4.0/}{\includegraphics[width=0.90\textwidth]{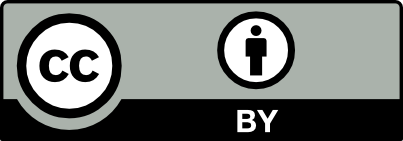}}
  \end{minipage}\hfill
  \begin{minipage}{0.7\columnwidth}
     \href{http://creativecommons.org/licenses/by/4.0/}{This work is licensed under a Creative Commons Attribution International 4.0 License.}
  \end{minipage}
  
  \vspace{5pt}
}{}{}

\makeatother

\begin{document}

%%
%% The "title" command has an optional parameter,
%% allowing the author to define a "short title" to be used in page headers.

\title{Answering Count Queries with Explanatory Evidence}

%%
%% The "author" command and its associated commands are used to define
%% the authors and their affiliations.
%% Of note is the shared affiliation of the first two authors, and the
%% "authornote" and "authornotemark" commands
%% used to denote shared contribution to the research.

\author{Shrestha Ghosh}
\affiliation{%
  \institution{Max Planck Institute for Informatics}
  \institution{Saarland University}
  \city{Saarbruecken}
  \country{Germany}
}
\email{ghoshs@mpi-inf.mpg.de}

\author{Simon Razniewski}
\affiliation{%
  \institution{Max Planck Institute for Informatics}
  \city{Saarbruecken}
  \country{Germany}
}
\email{srazniew@mpi-inf.mpg.de}

\author{Gerhard Weikum}
\affiliation{%
  \institution{Max Planck Institute for Informatics}
  \city{Saarbruecken}
  \country{Germany}
}
\email{weikum@mpi-inf.mpg.de}

% \authornotemark[1]

%% By default, the full list of authors will be used in the page
%% headers. Often, this list is too long, and will overlap
%% other information printed in the page headers. This command allows
%% the author to define a more concise list
%% of authors' names for this purpose.
\renewcommand{\shortauthors}{Ghosh, et al.}
%%
%% The abstract is a short summary of the work to be presented in the
%% article.
\begin{abstract}
A challenging case in web search and question answering are count queries, such as \textit{``number of songs by John Lennon''}. Prior methods merely answer these with a single, and sometimes puzzling number or return a ranked list of text snippets with different numbers. This paper proposes a methodology for answering count queries with inference, contextualization and explanatory evidence. Unlike previous systems, our method infers final answers from multiple observations, supports semantic qualifiers for the counts, and provides evidence by enumerating representative instances. Experiments with a wide variety of queries show the benefits of our method. To promote further research on this underexplored topic, we release an annotated dataset of 5k queries with 200k relevant text spans.
\end{abstract}

%%
%% The code below is generated by the tool at http://dl.acm.org/ccs.cfm.
%% Please copy and paste the code instead of the example below.
%%
% \begin{CCSXML}
% <ccs2012>
%  <concept>
%   <concept_id>10010520.10010553.10010562</concept_id>
%   <concept_desc>Computer systems organization~Embedded systems</concept_desc>
%   <concept_significance>500</concept_significance>
%  </concept>
%  <concept>
%   <concept_id>10010520.10010575.10010755</concept_id>
%   <concept_desc>Computer systems organization~Redundancy</concept_desc>
%   <concept_significance>300</concept_significance>
%  </concept>
%  <concept>
%   <concept_id>10010520.10010553.10010554</concept_id>
%   <concept_desc>Computer systems organization~Robotics</concept_desc>
%   <concept_significance>100</concept_significance>
%  </concept>
%  <concept>
%   <concept_id>10003033.10003083.10003095</concept_id>
%   <concept_desc>Networks~Network reliability</concept_desc>
%   <concept_significance>100</concept_significance>
%  </concept>
% </ccs2012>
% \end{CCSXML}

% \ccsdesc[500]{Computer systems organization~Embedded systems}
% \ccsdesc[300]{Computer systems organization~Redundancy}
% \ccsdesc{Computer systems organization~Robotics}
% \ccsdesc[100]{Networks~Network reliability}



%%
%% Keywords. The author(s) should pick words that accurately describe
%% the work being presented. Separate the keywords with commas.


\maketitle

% \begin{table}
%   \caption{Frequency of Special Characters}
%   \label{tab:freq}
%   \begin{tabular}{ccl}
%     \toprule
%     Non-English or Math&Frequency&Comments\\
%     \midrule
%     \O & 1 in 1,000& For Swedish names\\
%     $\pi$ & 1 in 5& Common in math\\
%     \$ & 4 in 5 & Used in business\\
%     $\Psi^2_1$ & 1 in 40,000& Unexplained usage\\
%   \bottomrule
% \end{tabular}
% \end{table}

% \begin{table*}
%   \caption{Some Typical Commands}
%   \label{tab:commands}
%   \begin{tabular}{ccl}
%     \toprule
%     Command &A Number & Comments\\
%     \midrule
%     \texttt{{\char'134}author} & 100& Author \\
%     \texttt{{\char'134}table}& 300 & For tables\\
%     \texttt{{\char'134}table*}& 400& For wider tables\\
%     \bottomrule
%   \end{tabular}
% \end{table*}

\section{Introduction}\label{sec:introduction}

% \vspace*{-0.1cm}
\paragraph{Motivation and Problem} 
Question answering (QA)
%(QA) and entity search with telegraphic queries over web contents
and web search with telegraphic queries
have been greatly advanced over the last decade \cite{DBLP:series/irs/Balog18,DBLP:journals/kais/DiefenbachLSM18,DBLP:journals/access/HuangXHWQFZPW20,DBLP:journals/semweb/UsbeckRHCHNDU19}.
Nevertheless, queries that can have multiple correct answers due to variance in semantic qualifiers (``top 10 albums'', ``singles albums'', ``remastered albums'') and alternative representations through instances remain underexplored and pose open challenges.
% Nevertheless, specific classes of queries are underexplored and pose open challenges.
This paper addresses the class of \textit{count queries},
to return the number of instances that have a certain property. Examples 
are:

% \vspace*{0.1cm}
\begin{itemize}
%\medskip
\item \textit{How many songs did John Lennon write for the Beatles?}
% \item \textit{How many people won the Fifa world cup both as player and as manager?}
\item \textit{How many languages are spoken in Indonesia?}
\item \textit{How many unicorn companies are there?}
%\medskip
\end{itemize}

Count queries are frequent in search engine logs as well as QA benchmarks 
~\cite{rajpurkar2016squad,kwiatkowski2019natural,dubey2019lc,voorhees2001overview}.
If the required data is in a structured knowledge base (KB) such as Wikidata~\cite{vrandevcic2014wikidata}, then answering is relatively straightforward.
However, KBs are limited not only by their sparsity, but also by the lack of direct links between instances and explicit counts when both are present. 
Besides, evaluating the additional condition \textit{``for the Beatles''} (i.e., a subset of his songs) is beyond their scope. 

Search engines handle popular cases reasonably well, but also
fail on semantically refined requests (e.g., \textit{``for the Beatles''}), merely returning either a number without explanatory evidence or multiple candidate answers with high variance. 
Answering count queries from web contents thus poses several challenges:

\begin{enumerate}
    \item[1.] \textit{Aggregation and inference:} 
    Returning just a single number from the highest-ranked page can easily go wrong.
    Instead, joint inference over a set of candidates, with an awareness of the distribution and other signals, is necessary for a high-confidence answer.
    \item[2.] \textit{Contextualization:} Counts in texts often 
    come with contexts on the relevant instance set.
    % , such as ``in high school'', or ``for the Beatles''. 
    For example, John Lennon co-wrote about 180 songs 
    %together with Paul McCartney, 
    for the Beatles,
    % 43 credited to Lennon-McCartney on his own, 
    150 as a solo artist, etc.
    For correct answers it is crucial
    to 
    %go beyond IR-style context similarity, and to explicitly extract and process these semantic qualifiers.
    capture context from the underlying web pages
    and properly evaluate these kinds of semantic qualifiers.
    \item[3.] {\textit{Explanatory Evidence:}} 
    A numeric answer alone, such as \textit{``180''} for the Beatles songs by Lennon, is often unsatisfactory. 
    The user may even perceive this as non-credible, and think that it is too high as she may have only popular songs in mind.
    %thinking only in terms of the solo authored ones, which are much less. 
    It is, therefore, crucial 
    % provide user-comprehensible explanations, 
    %{explain these counts,}
    %by showing notable instances.
    to provide users with explanatory evidence.
    % and web evidence for them.
\end{enumerate}

\paragraph{Approach and Contribution} 
This paper presents {\mysystem}, \underline{Co}unt \underline{Q}uestion answering with \underline{Ex}planatory evidence, which answers count queries via three components: i) \textit{answer inference} ii) \textit{answer contextualization} and, iii) \textit{answer explanation}.

Given a full-fledged question or telegraphic query and relevant text passages, {\mysystem} 
% retrieves a set of candidate results and
applies joint inference to compute a high-confidence answer for the count itself.
It provides contextualization of the returned count answer,
through semantic qualifiers into equivalent or subclass categories, and extracts a set of representative instances as explanatory evidence, exemplifying the returned number for
enhanced credibility and user comprehension.
%retrieves a ranked-list of instances that explain the counts.
%
% , by using ranking techniques to identify informative instances.
% The results from both tasks are combined to boost the accuracy of each other. 

\noindent Novel contributions of this work are:

\begin{enumerate}
    \item[1.] introducing the problem of 
    count query answering with explanatory evidence;
    \item[2.] developing a 
    method
    %module 
    for inferring high-confidence counts from noisy candidate sets;
    \item[3.] developing 
    %two modules 
    techniques
    to provide answer contextualization and explanations;
    % from count-modified noun phrases, %%SG: I'm refraining from using the term CNPs until introduced in CoRe in the methodology.
    \item[4.] evaluating 
    % Co$^3$ 
    {\mysystem}
    against state-of-the-art baselines 
    %and diverse query datasets, and; 
    on a variety of test queries;
    % and
    \item[5.] releasing an annotated data resource with 5k count queries and 200k text passages,
    % with annotations,
    with preliminary access for reviewing at {\small\url{\systemurl}}.
    % {\small\url{https://tinyurl.com/countqueryappendix}}.
\end{enumerate}

\section{Related Work}

Where structured data is available in KBs, structured QA is the method of choice.
%, and previous work has looked at identifying count information inside KBs~\cite{ghosh2020uncovering} 
 However, for many topics, no relevant count information can be found in KBs. 
For example, Wikidata contains 217 songs attributed to John Lennon\footnote{\url{https://w.wiki/4XVq}},

but is incomplete in indicating whether these written for the Beatles or otherwise. {Previous analyses found that 5-10\% of queries in popular QA datasets are count queries~\cite{enriching2018mirza}}.
In the KB-QA domain, systems like QAnswer~\cite{diefenbach2019qanswer} tackle count queries
by aggregating instances using the SPARQL \texttt{count} modifier. This is liable to incorrect answers, when instance relations are incomplete.
% Attempts have also been made to improve recall by hybrid QA over text and KB, yet without specific consideration of counts~\cite{lu2019answering,xu2016hybrid}.
{Recent research has also explored QA over web tables, with a focus on table retrieval and span prediction~\cite{herzig2021open}. Another challenging aspect of KB-QA is reducing the semantic gap between natural language query and the SPARQL query formulation~\cite{berant2013semantic}, which would in turn lead to better count query answering.}

% \paragraph{Count Information in QA} 
% In~\cite{enriching2018mirza}, it is reported that 5\%-10\% of queries in popular QA datasets are count queries. 
State-of-the-art systems typically approach QA via the machine reading paradigm~\cite{karpukhin2020dense,joshi2020SpanBERT,sanh2019distilbert,chen2017reading,dua2019drop}, where the systems find the best answer in a given passage. The retriever-reader approach in open-domain QA uses several text segments 
% (multi-paragraphs from a single or multiple documents) 
to return either a single best answer~\cite{chen2017reading,wang2018evidence} or a ranked list of documents with the best answer per document~\cite{karpukhin2020dense}.
The DPR system~\cite{karpukhin2020dense}\footnote{\url{http://qa.cs.washington.edu:2020}} returns 
\textit{``approximately 180''} from its rank-1 text passage
to both, the simple John Lennon query, and the refined variant with \textit{``\dots for the Beatles''}. 
The other top-10 snippets include false results such as \textit{``five''} and contradictory information such as \textit{``180 jointly credited''}
% Lennon and McCartney together, 
(as if Lennon had not written any songs alone).
Thus, QA systems are not robust (yet) and lack explanatory evidence beyond merely returning the top-ranked text snippet. 

Attempts have also been made to improve recall by hybrid QA over text and KB, yet without specific consideration of counts~\cite{lu2019answering,xu2016hybrid}.
Search engines can answer simple count queries from their underlying KBs, if present, a trait which we exploit to create our CoQuAD dataset (Section.~\ref{sec:dataset}). But more often they return informative text snippets, similar to QA-over-text systems.
The basic Lennon query has a highest-ranked Google snippet with
\textit{``more than 150''} when given the telegraphic input
\textit{``number of songs by John Lennon''} and
\textit{``almost 200''} when given the full-fledged question
\textit{``how many songs did John Lennon write''}. For the latter case,
the top-ranked snippet talks about the composer duo
\textit{``John Lennon and Paul McCartney''}. 
When refining the query by qualifiers, this already puzzling
situation becomes even more complex with \textit{``84.55 of 209 songs''} being ranked first followed by varying counts such as \textit{``18 Beatles songs''} 
(co-written with McCartney) 
and \textit{``61''} 
(written separately).  
Because of the lack of consolidation, the onus is on the user to decide whether there are multiple correct answers across text segments.

% \paragraph{Long Form Question Answering}
It is recognized that just literally answering questions is often not sufficient for use cases. One line of work tackles this by returning comprehensive answer in full sentences, using templates~\cite{kacupaj2020vquanda}. Another line concerns long form question answering, where the QA model retrieves multiple relevant documents to generate a whole answer paragraph~\cite{krishna2021hurdles}. The ELI5 dataset~\cite{fan2019eli5} contains diverse open-ended queries with supporting information from relevant web sources. While the setting is related, long form QA is concerned with generating textual answers evidenced on multiple documents, while we focus on answering count queries by consolidating counts and grounding them in instances.

\begin{figure}[t]
    \centering
    \includegraphics[width=0.48\textwidth,trim={3.8cm 2.3cm 3.5cm 2.3cm},clip]{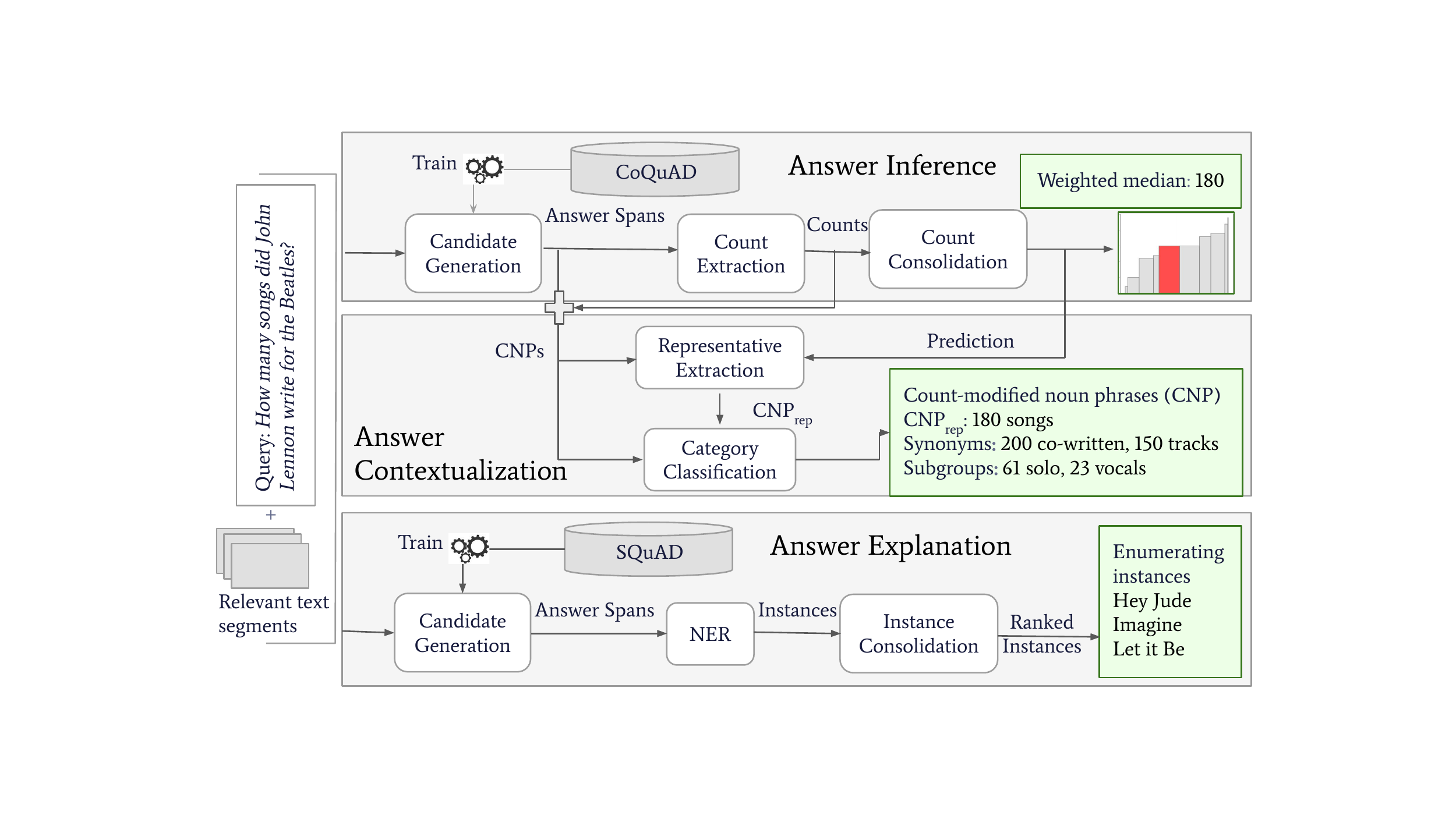}
    \caption{System overview of {\mysystem}.}
    \label{fig:system_arch}
    % \vspace{-0.6cm}
\end{figure}

\section{Methodology}

We approach count question answering by a combination of per-document answer span prediction, context extraction, and consolidation of counts and instances across documents.
%For count queries, answer types are uniformly counts, and the mileage thus comes from recognizing counts in the right context, and consolidating corroborating candidate answers into a most likely value.
% We therefore approach the task of count QA over text by utilizing evidence from multiple document sources for solving the tasks of \textit{answer inference}, \textit{answer contextualization} and 
% \textit{count explanation}
Fig.~\ref{fig:system_arch} gives the overview of {\mysystem}. We consider as input a query that asks for the count of named entities that stand in relation with a subject, for instance full queries like \textit{``How many songs did John Lennon write for the Beatles''}, or a keyword query like \textit{``songs by lennon''}.

We further assume that relevant documents or passages are given. This could be the result of a standard keyword/neural embedding-based IR procedure over a larger (locally indexed) background corpus, like Wiki\-pedia or the Web. % Yet this is merely a pragmatic choice, locally indexed text corpora could be plugged in as well.
We focus on extracting counts and instances (entity-mentions) from the text segments so as to i) consolidate the counts to present the best answer, ii) present 
contextualization
as a means to semantically qualifying the predicted count, and iii) ground the count in instances.

\paragraph{Answer Inference}\label{subsec:coin}
For obtaining answer candidates, we use the popular SpanBERT model~\cite{joshi2020SpanBERT}, trained on the CoQuAD train split for candidate generation.
Span prediction models 
return general text spans, which may contain worded answers (\textit{five children}, $\textit{Conf}=0.8$), modifier words and other context 
(\textit{17 regional languages}, $\textit{Conf}=0.75$). 
These answer spans have two components - the count itself and qualifiers, which we separate with the help of the CogComp Quantifier~\cite{roy2015reasoning}.
To consolidate the resulting candidate counts, we compare four methods:
% \begin{enumerate}
\squishlist
    \item[1.] \textit{Most confident:} The candidate given the highest confidence by the neural model. This is commonly used in textual QA~\cite{chen2017reading,wang2018evidence}.
    
    \item[2.] \textit{Most frequent:} A natural alternative is to rank answers by frequency, and prefer the ones returned most often. 
    
\squishend
While \textit{most confident} may be susceptible to single outliers, \textit{most frequent} breaks down in cases where there are few answer candidates. But unlike textual answers, numbers allow further statistical aggregation:
% \begin{enumerate}[resume]
\squishlist
    \item[3.] \textit{Median:} The midpoint in the ordered list of candidates.
    \item[4.] \textit{Weighted Median:} The median can be further adapted by weighing each candidate with the model's score.
\squishend

\paragraph{Answer Contextualization}
% \subsubsection{\bf Count-modified Noun Phrases}
The answer candidates from the previous module often contain nouns with phrasal modifiers, such as \textit{17 regional languages}. %The answer span, count and score tuples we obtained in CoIn are re-used in CoRe. Not all answer spans contain counts (``few languages'', $\textit{count}=0$, $\textit{score}=0.5$). To those that do contain them we define as 
We call these 
\textit{count-modified noun phrases} (CNPs). These CNPs stand in some relation with the predicted count from the answer inference module. 
The representative CNP, CNP$_\text{rep}$,  
% As representative CNP, we use 
which best accompanies the predicted count is first chosen and then compared with the remaining CNPs. Since answer inference uses a consolidation strategy, we select the CNP with count within $\pm\alpha$ of the predicted count having the highest confidence as CNP$_\text{rep}$, where $\alpha$ is between 0 and 100\%, 0 being most restrictive.
The remaining CNPs are categorized as follows:
% By comparing count-modified noun phrases (CNPs) based on their relation with the representative CNP of the predicted answer (CNP$_\textit{rep}$), we can discern their semantics:
\begin{enumerate}
    \item[1.] \textit{Subgroups}: CNPs which are semantically more specific than CNP$_\textit{rep}$, and are expected to count only a subset of the instances counted by CNP$_\textit{rep}$. %  where the accompanying count is lower than the predicted answer. E.g., \textit{regional languages} is predicted as subgroup of \textit{language}, if it occurs with counts $\langle 23, 17, 42\rangle$ and the predicted query answer is 700.
    \item[2.] \textit{Synonyms}: CNPs, whose meaning is highly similar to CNP$_\textit{rep}$. %accompanying count is within a specified threshold of the predicted count. E.g., \textit{tongue} is predicted as synonym of \textit{language}, if it occurs with counts $\langle 530, 810, 600\rangle$.
    \item[3.] \textit{Incomparables}: CNPs which count instances of a completely different type.
    %If a CNP is semantically dissimilar to CNP$_\textit{rep}$ or the accompanying count of CNP is much greater than the predicted count we categorize it as unrelated. E.g., \textit{speakers}, which occurs with counts in the order of millions, is semantically similar to \textit{language}, but does not relate to the language count.
\end{enumerate}

\noindent
We assign these categories based on (textual) semantic relatedness of the phrasal modifier, and numeric proximity of the count.
For example, \textit{regional languages} is likely a subgroup of \textit{700 languages}, especially if it occurs with counts $\langle 23, 17, 42\rangle$.
\textit{tongue} is likely a synonym, especially if it occurs with counts $\langle 530, 810, 600\rangle$. \textit{Speakers} is most likely incomparable, especially if it co-occurs with counts in the millions.

CNPs with embedding-cosine similarity~\cite{reimers2019sentencebert} less than zero are categorized as incomparable, while from the remainder, those with a count within $\pm\alpha$ are considered synonyms, lower count CNPs are categorized as subgroups, and higher count CNPs as incomparable. 

For instance, for the query ``How many languages are spoken in Indonesia'', with a prediction 700, \textit{estimated 700 languages} would be the $\textit{CNP}_\textit{rep}$, \{\textit{700 languages}, \textit{750 dialects}\} would be classified as synonyms, \{\textit{27 major regional languages}, \textit{5 official languages}\} as subgroups and \{\textit{2000 ethnic groups}, \textit{85 million native speakers}\} as incomparables.

\paragraph{Answer Explanation}\label{subsec:coex}
% \paragraph{Enumerating Instances}
Beyond classifying count answer contexts, showing relevant sample instances is an important step towards explainability. To this end, we aim to identify entities that are among the ones counted in the query.

We again use the SpanBERT model to obtain candidates, this time with a modified query, replacing "how many" in the query with "which" (or adding it), so as to not confuse the model on the answer type. We extract named entities from the answer spans and rank them
% We now derive ranked lists of these instances 
using the following alternative approaches:
\begin{enumerate}
% \squishlist
\item[1.] \textit{QA w/o Consolidation.} In the spirit of conventional QA, where results come from a single document, we return instances from the document with the most confident answer span.

\item[2.] \textit{QA + Context Frequency.} 
The instances are ranked by their frequency.%, $score_i = \frac{|D_i|}{|D|}$.

\item[3.] \textit{QA + Summed Confidence.} We rank the instances based on the summed confidence of all answer spans that contain them.%the parent answer span as returned by the QA model, $score_i = \sum_{d \in D_i} \text{Conf}_i^d$.

\item[4.] \textit{QA + Type Compatibility.} Here instances are ranked by their compatibility with the query's answer type, extracted via the dependency parse tree. We form a hypothesis \textit{``(instance) is a (answer type)''} and use the probability of its entailment from the parent sentence in the context from which the instance was extracted to measure type compatibility. We use~\cite{liu2019roberta} to obtain entailment scores, which are again summed over all containing answer spans. 

\end{enumerate}

\section{The C\lowercase{o}Q\lowercase{u}AD Dataset}\label{sec:dataset}

\paragraph{Dataset construction}
Existing QA datasets only incidentally contain count queries; we leverage search engine autocomplete suggestions to automatically compile count queries that reflect real user queries~\cite{how-google-autocomplete-predictions-work}. We provide the Google search engine with iterative query prefixes of the form \textit{``How many \texttt{X}''}, where $X \in \{a, aa, \ldots, zzz\}$, similar to the candidate generation from patterns used in~\cite{romero2019commonsense}, and collect all autocomplete suggestions via SERP API\footnote{\url{https://serpapi.com}}. We keep those with at least one named-entity (to avoid too general queries) and no measurement term (to avoid non-entity answer types). This gives us 11.3k count queries. 

We automatically obtain \textit{count ground truth} by collecting structured answers from the same search engine. Executing each query on Google, we scrape knowledge graph (KG) answers and featured snippets, using an off-the-shelf QA extraction model~\cite{sanh2019distilbert} to obtain best answers from the latter. This gives us KG ground truth for 131 queries, and ground truth from featured snippets for 6.7k queries. We again discard queries whose automated ground truth answer contains a measurement term, and manually annotate 100 queries from those without automated ground truth.

We next scrape the top-50 snippets per query from Bing, and obtain 
\textit{text segment ground truth} by labelling answer spans returned by the count extractor~\cite{roy2015reasoning} as positive when the count lies within $\pm 10\%$ from the ground truth. 
There are around 800 queries with no positive snippets, which we do not discard, so the system is not forced to generate an answer. In the end we have $5162$ count queries with automated ground truth, and an average of $40$ annotated text segments per query. 
We use 80\% of the count queries with automated ground truth for training. The test set contains 50 count queries with KG ground truth, 100 with ground truths from featured snippets, and 100 with manually annotated ground truth for quantitative and qualitative analysis.
We also manually annotate 75 queries with at least top-5 prominent instances for evaluating answer explanations.

\paragraph{Dataset Characteristics}
Queries in CoQuAD cover a range of topics, notably entertainment (27\%), social topics (20\%), organizations (12\%), technology (8\%) and politics (7\%). We find that 20\% of query results are fully stable (a company's founders, casts in produced movies), 55\% are low-volatile (lakes in a region, band members of an established but active band), 25\% are near-continuous (employment numbers, COVID cases). 
Most queries count entities in a simple relation to one named entity, i.e., the avg.\ query length is 6.40 words, with an average of 1.08 named entities per query.

\smallskip

\noindent Research access to the data is at {\small\url{https://tinyurl.com/countqueryappendix}}.

\section{Evaluation}
\paragraph{Implementation Details}
{The transformer model for answer inference candidate generation is trained for 2 epochs, at a learning rate of $3e^{-5}$. An input datapoint consists of a query, a text segment and a text span containing the count answer (empty if no answer). We train over 3 seeds and report the average score. For getting the instances from the answer spans, we use the pre-trained SpaCy NER model\footnote{\url{https://spacy.io} on the \texttt{en\_core\_web\_sm} model.}.}

\paragraph{Component analysis of {\mysystem}} 
We evaluate the {\mysystem} components to determine the best configurations for answer inference, consolidation and explanation. While we can use regular IR metrics of precision and recall for evaluating answer explanations (Precision@k, Recall@k, Hit@k and MRR) and accuracy of classification for evaluating count context categories, we need a new metric for counts. We use \textit{Relaxed Precision} (RP), which is the fraction of queries where the prediction lies within $\pm10\%$ of the gold answer. We also report \textit{Coverage}, which measures the fraction of queries that a systems returns an answer for.

We test the candidate generator for count spans on SpanBERT~\cite{joshi2020SpanBERT} finetuned on
i)  CoQuAD and,
ii) the popular general QA dataset SQuAD~\cite{rajpurkar2016squad}.
\textit{Fine-tuning} on SQuAD gives slightly higher precision scores (29.8\% vs.\ 27.4\% RP), but CoQuAD gives a higher coverage (82\% vs.\ 73.8\%) resulting in overall more correctly answered queries. 

For \textit{answer inference}, the \textit{frequent} and \textit{weighted median} consolidation schemes outperform the others, with \textit{weighted median} achieving 27.4\% RP just 0.6\% ahead of \textit{frequent}. Thus, for queries backed by less variant data, \textit{frequent} is good enough, but to have an edge in more variant data \textit{weighted median} is the way to go.

We assess the classification accuracy of CNPs for a manually labelled sample of 294 CNPs for 64 queries. While a strict threshold of 10\% ensures a high accuracy of 90\% for \textit{Synonyms}, which only decreases with increasing $\alpha$, the accuracy of \textit{Subgroups} and \textit{Incomparables} is initially low (at or less than 50\%), and peaks only at a much higher $\alpha$. A weighted optimum is reached at $\alpha=30\%$, where the accuracy of the \textit{Synonyms} does not degrade much (89\%), and the accuracy of \textit{Subgroups} and \textit{Incomparables} is both above 60\% (61\% and 62\% respectively).

In \textit{answer explanation}, \textit{QA + context frequency} consolidation performs consistently well. While \textit{QA w/o consolidation} has highest P (11.5\%) and R (2.3\%) at rank 1, its performance decreases subsequently to only 2.7\% P and 5.1\% R at rank 10, compared to 10.8\% P and 18.2\% R of \textit{QA + context frequency}. This indicates that the QA model is tailored to the typical setting QA of a single correct answer, and that consolidation helps beyond that.

\paragraph{Comparing CoEx with baselines}
We compare our proposed system with two complementary paradigms.
\begin{enumerate}
    \item[1.] Knowledge-base question answering: QAnswer~\cite{diefenbach2019qanswer}.
    \item[2.] Commercial search engine QA: Google Search Direct Answers (GoogleSDA).
\end{enumerate}
%
% \noindent\textbf{Q5. How well does {\mysystem} answer count queries?}\newline
For fairness to QAnswer, we
%, which specifically deals with count queries by aggregating on top of the SPARQL query, we queried the system\footnote{QAnswer API at \url{https://qanswer-core1.univ-st-etienne.fr/swagger-ui.html}} twice - the original count query (for the count answer) and a modified variant as in Section~\ref{subsec:coex}, i.e., replacing \textit{``how many''} with \textit{``which''}. We then
post-processed the results to extract count and instances.
% Secondly, we post-processed answers, applying additional count aggregation on QAnswer results that were sets, and building scrapers for GoogleSDA answers that came from featured snippets, to increase its recall beyond direct KB answers. %. We only excluded the case of text snippets - considering those would significantly boost GoogleSDA's coverage on Stresstest from 37\% to 60\%, the catch being that the text would require significant post-processing for answer extraction. 
For evaluating instances by GoogleSDA, we post-processed knowledge graph and featured snippet of the search engine result page, keeping items from list-like structures as instances ranked in their order of appearance. %All evaluation is done on 250 manually annotated count queries.

\begin{table}[t]
    \caption{Comparing answer inference results (in percentages).}
    \label{tab:different_datasets}
    % \vspace{-0.4cm}
    \begin{adjustbox}{width=0.35\textwidth,center}
    \centering
    \begin{tabular}{|l|ll|ll|ll|}
    \hline
        \multirow{2}{1cm}{\textbf{System}} & \multicolumn{2}{c|}{\textbf{CoQuAD}} & \multicolumn{2}{c|}{\textbf{LCQuAD$_\textit{count}$}} & \multicolumn{2}{c|}{\textbf{Stresstest}}\\
        & \textbf{RP} & \textbf{Cov} & \textbf{RP} & \textbf{Cov} & \textbf{RP} & \textbf{Cov} \\
        \hline
        QAnswer~\cite{diefenbach2019qanswer} &  6.6 & \textbf{95.6} &
        \textbf{29.8} &  \textbf{50.0} & 3.0 & 45.0\\
        GoogleSDA & \textbf{84.6} & 60.0 & 5.7 & 10.5 & 22.0 & 37.0 \\
        \hline
        {\mysystem} & 27.4 & 82.0 & 6.7 &  45.2 & \textbf{38.6} & \textbf{88.3} \\
        \hline 
    \end{tabular}
    \end{adjustbox}
    
\end{table}

\begin{table}[t]
    \caption{Precision@k (P@k), Recall@k (R@k), Hit@10 and MRR for the answer explanations of {\mysystem} and baselines.
    }
    \label{tab:enum_comparative}
    % \vspace{-0.4cm}
    \begin{adjustbox}{width=0.45\textwidth,center}
    \centering
    \begin{tabular}{|l|lll|lll|l|l|}
    \hline
        % \multirow{2}{1cm}
        {\textbf{System}}
        & \textbf{P@1} & \textbf{P@5} & \textbf{P@10}
        & \textbf{R@1} & \textbf{R@5} & \textbf{R@10}
        & \textbf{Hit@10} 
        & \textbf{MRR}\\
        \hline
        QAnswer~\cite{diefenbach2019qanswer} & 4.7 & 4.5 & 3.7 
        & \textbf{2.1} & 0.8 & 1.1
        & 12.8 
        & 0.064\\
        GoogleSDA & \textbf{7.7} & \textbf{14.2} & \textbf{21.7} 
        & 1.3 & 2.2 & 3.2
        & 12.8 
        & 0.089\\
        \hline
        \mysystem & {6.4} & {12.8} & {10.8}
        & {1.3} & \textbf{11.7} & \textbf{18.2} 
        & \textbf{55.1}
        & \textbf{0.218} \\ 
        \hline 
    \end{tabular}
    \end{adjustbox}
\end{table}

For \textit{answer inference}, Table~\ref{tab:different_datasets} compares the three systems on the 250 annotated test CoQuAD queries. We also present the performances on 100 count queries from an existing dataset LCQuAD~\cite{dubey2019lc}, and a manually curated dataset of 100 challenging count queries called \textit{Stresstest}. While GoogleSDA has a high precision on CoQuAD (consisting 150 KG and snippet answerable queries), {\mysystem} not only provides high coverage but a decent RP. On LCQuAD, a dataset designed specifically for KG queries, {\mysystem} loses to LCQuAD while still maintaining a better coverage and RP score compared to GoogleSDA. The results indicate that reliance on structured KBs (QAnswer) is not sufficient for general queries, and robust consolidation
from crisper text segments is necessary.

For \textit{answer explanation}, the comparison results on the 75 instance-annotated CoQuAD queries are in Table~\ref{tab:enum_comparative}. {\mysystem} provides more R than the baselines, with competitive P at rank 1, and better at ranks 5 and 10. {\mysystem} performs consistently better at hits@k and MRR, losing only slightly in precision at ranks 1 and 5. Both baselines answer less than 25 queries at rank 5, and at rank 10 less than 20 queries, and QAnswer performs poorly in the returned answers. GoogleSDA operates extremely conservative, thus maintaining precision at lower ranks, at the cost of tiny recall.

\smallskip

\begin{table}[t]
    \caption{Extrinsic user study (precision in percent).}
    \label{tab:user_study_eval}
    % \vspace{-0.4cm}
    \begin{adjustbox}{width=0.43\textwidth,center}
    \centering
    \begin{tabular}{|l|lllll|}
    \hline
    \textbf{Class} & \textbf{Only Count} & \textbf{+Instances} & \textbf{+CNPs} & \textbf{+Snippet} & \textbf{All} \\
    \hline
    Correct & 73 & 63 & 78 & 75 & \textbf{88}\\
    Incorrect & 28 & 45 & 40 & \textbf{53} & 45\\
    Both & 55 & 56 & 63 & 66 & \textbf{71} \\
   \hline
    \end{tabular}
    \end{adjustbox}
    % \vspace{-0.5cm}
\end{table}

\paragraph{Qualitative Comparison}
We looked further into the 250 queries (50 KG answerable, 100 snippet answerable and 100 with ranked snippets) for understanding how baselines tackle progressively difficult queries. 
As explained in the CoQuAD dataset creation, we encountered three ways in which Google answers count queries. Around 1\% of the answers came from the Google KG, 59\% through featured snippets, 40\% through page results. Besides these high-level categories, answers can be categorized by the following aspects:
\begin{enumerate}
\item[1.] Listing only instances (e.g., for ``mayors of New York''),
\item[2.] Listing only counts (e.g., for ``employees of NHS''),
\item[3.] Listing both instances and counts, or
\item[4.] Listing counts refined with semantic qualifiers (e.g., ``7 official languages and 30 regional languagues'').
\end{enumerate}
% We look for such cues in our evaluation datasets (see Table~\ref{tab:count_answer_availability_type}).
On the above 250 queries, 
% addressed simon's comment on specifying which data
QAnswer returns mostly counts (95.6\%) and rarely instances (2\%) or no answers (2.4\%). Among the KG-based answers, GoogleSDA returns both counts and instances for 90\% of the queries, and only counts for the remaining 10\%. Among the featured-snippets-based answers, 85\% contain only counts, and just 15\% contain both counts and instances. Semantic qualifiers are common in featured snippets, coming up for 73\% of queries. While semantic qualifiers can be expressed in KG answers (``volcanic islands in Hawaii''$\Rightarrow$islands$\rightarrow$Hawaii$\rightarrow$volcanoes), this rarely shows up, unless the queried entity and the qualifier are extremely popular.

\paragraph{User Studies}
We asked 120 MTurk users for pairwise preferences between answer pages that reported bare counts, and counts enhanced by either of the explanation types. 50\% of participants preferred interfaces with CNPs, 80\% with a snippet, 73\% with instances, 63\% preferred an interface with all three enabled (remaining percentage: prefer bare count answer/same/cannot decide). While snippets are already in use in search engines, the results indicate that CNPs and instances are considered valuable, too.

We also validated the merit of explanations extrinsically. We took 5 queries with correct count results, 5 with incorrect results, and presented the system output under the 5 explanation settings to 500 users. The users' task was to judge the count as correct or not based on the explanations present. The measured precision scores are in Table~\ref{tab:user_study_eval}. All explanation had a positive effect on overall annotator precision, especially for incorrect counts.

\section{Conclusion}

{In this work we highlighted the distinct challenges of count query answering, in particular, that count answer candidates form numeric distributions, that answer contexts stand in hierarchical relations, and that counts coexist with instances, that ideally should be returned as well.}
We introduced an approach for count query answering that contains these three components.

{We construct the CoQEx dataset of count queries that reflect real user needs. Through {\mysystem}, we show how to provide a better trade-off between precision and coverage than the current deployed systems, on multiple challenging datasets. User studies show that explanatory evidence improves user comprehension, especially through CNPs when there exist similar competing answers attributed to specific qualifiers.}
To foster further research, we release all 
datasets\footnote{\small\url{\systemurl}}.

%%
%% The acknowledgments section is defined using the "acks" environment
%% (and NOT an unnumbered section). This ensures the proper
%% identification of the section in the article metadata, and the
%% consistent spelling of the heading.
\begin{acks}
We would like to thank the anonymous reviewers for their valuable feedback and suggestions.
% To Robert, for the bagels and explaining CMYK and color spaces.
\end{acks}

%%
%% The next two lines define the bibliography style to be used, and
%% the bibliography file.
\bibliographystyle{ACM-Reference-Format}
\bibliography{refs}

%%
%% If your work has an appendix, this is the place to put it.
% \appendix
\end{document}